\documentclass[journal]{IEEEtran}
\usepackage{amsmath,amsthm,amssymb}
\usepackage{blkarray}
\usepackage{subcaption}
\usepackage{graphicx}
\usepackage{multirow}
\usepackage{tabularx}
\usepackage{colortbl}
\usepackage{newalg}
\usepackage{tikz}

\usepackage[inline, shortlabels]{enumitem}

\makeatletter
 \newcommand*{\@rowstyle}{}
\newcommand*{\rowstyle}[1]{
 \gdef\@rowstyle{#1}%
 \@rowstyle\ignorespaces%
}

\DeclareMathOperator{\F}{\mathbb F}

\theoremstyle{plain}

\newcommand{\mF}{\mathcal F}

\newcommand{\set}[1]{\left\{{#1}\right\}}

\newcommand{\E}[2]{\mathbf E_{{#1}}\left[{#2}\right]}

\DeclareMathOperator{\sgn}{sgn}

\begin{document}

\title{Performance and Complexity of the Sequential Successive Cancellation Decoding Algorithm}
\author{Peter Trifonov, ~\IEEEmembership{Member, IEEE} 
}

\maketitle

\begin{abstract}
Simulation results illustrating the performance and complexity of the sequential successive cancellation decoding algorithm are presented for the case of polar subcodes with Arikan and large kernels, as well as for extended BCH\ codes. Performance comparison with Arikan PAC and LDPC codes is provided.
Furthermore, complete description of the decoding algorithm is presented.
\end{abstract}
\section{Introduction}
Polar codes are a novel class of error correcting codes, which was already adopted for use in 5G systems \cite{arikan2009channel}.  The development of polar codes started from the analysis of sequential decoding \cite{arikan2016origin}.  It appears that the sequential decoding techniques can be applied to decoder polar codes as well \cite{niu2012stack,trifonov2013polar,miloslavskaya2014sequential,trifonov2018score,miloslavskaya2014sequentialBCH}. In this memo we present some performance and complexity results for the sequential decoding algorithm introduced in \cite{miloslavskaya2014sequential} and later refined in \cite{trifonov2018score}.   Furthermore, we present a complete description of this algorithm. Although the algorithm was originally introduced in the context of Arikan polar codes, it can be used for decoding of polar codes with large kernels.  The algorithm can also naturally handle codes with dynamic frozen symbols, so it can be used for decoding of polar subcodes and extended BCH\ codes.  

\section{Background}
 An $(n = l^m, k)$ polar code is a linear block code generated by $k$ rows of matrix $G_m = K^{\otimes m}$, where $\otimes m$ is $m$-fold Kronecker product of matrix with itself, $K$ is a kernel, and  $[l]=\set{0,\dots,l-1}$.
The encoding  scheme is given by
$ c_0^{n-1}= u_0^{n-1}G_m$,
where $u_i,i\in \mathcal F$ are  set to some pre-defined values, e.g. zero (frozen symbols),  $|\mF| = n - k$, and the remaining values $u_i$ are set to the payload data. 

More general construction is obtained by allowing some frozen symbols to be equal to linear combinations of other symbols, i.e. 
\begin{equation}
\label{mDynFrozen}
u_{i_j}=\sum_{s=0}^{i_j-1}u_sV_{js}, 0\leq j<n-k,
\end{equation}
where $V$ is a constraint matrix, such that last non-zero elements of its rows are located in distinct columns $i_j$.   The constraint matrix can be constructed using both algebraic and randomized techniques \cite{trifonov2016polar,trifonov2017randomized,trifonov2019construction,trifonov2020randomized}.
The symbols, where the right hand side of \eqref{mDynFrozen} is non-zero are also known as dynamic frozen \cite{trifonov2013polar} or  parity check frozen symbols  \cite{wang2016paritycheckconcatenated}.

In particular, the constraint matrix can be obtained as 
\begin{equation}
\label{mCheckConstraint}
V=QHG_m^T,
\end{equation}
 where $H$ is a check matrix of some linear block code of length $l^m$, and $Q$ is a suitable invertible matrix. This allows one to apply the techniques developed for decoding of polar codes  to other codes. Extended primitive narrow sense BCH\ codes were shown to be particularly well suited for this approach \cite{trifonov2016polar}.

Decoding of polar (sub)codes can be implemented by the successive cancellation (SC) algorithm.
It is  convenient to describe the SC algorithm in terms of probabilities $W_{m}^{(i)}\set{U_0^i=v_0^{i}|\mathbf Y=y_0^{n-1}}=W_{m}^{(i)}\set{v_0^{i}|y_0^{n-1}}$   of transmission of various vectors $v_0^{n-1}G_m$ with given values $v_0^i$, provided that the receiver observes a noisy vector $y_0^{n-1}$, i.e.
\begin{align}
\lefteqn{W_{m}^{(i)}\set{v_0^{i}|y_0^{n-1}}=\frac{W_m^{(i)}(y_0^{n-1},v_0^{i-1}|v_i)}{2W(y_0^{n-1})}}\nonumber\\
=&\sum_{v_{i+1}^{n-1}}W_{m}^{(n-1)}\set{v_0^{n-1}|y_0^{n-1}}
=\sum_{v_{i+1}^{n-1}}\prod_{j=0}^{n-1}W\set{(v_0^{n-1}G_{m})_j|y_j},
\label{mTotalProb}
\end{align}
where $W_m^{(i)}(y_0^{n-1},v_0^{i-1}|v_i)$ is the transition probability function for the $i$-th subchannel induced by the polarizing transformation $G_m$.

 At phase $i$ the SC decoder  makes decision 
\begin{equation*}
\widehat{u}_i=\begin{cases} \arg \max_{v_i\in\F_2} W_{m}^{(i)}\set{\widehat u_0^{i-1}.v_i|y_0^{n-1}},& i\not \in \mathcal F \\\sum_{s=0}^{i_j-1}\hat u_sV_{js},&\text{otherwise},\end{cases}
\end{equation*}
where $a.b$ denotes a vector obtained by appending $b$ to $a$.  

The SC decoder is known to be highly suboptimal. Much better performance can be obtained with the successive cancellation list (SCL)\ algorithm \cite{tal2015list}, which considers at each phase  $i$ at most   $L$ most probable paths $v_0^i$ satisfying freezing constraints. 
\section{Sequential decoding}
In this section we review of the sequential decoding algorithm introduced in \cite{trifonov2018score}.

The  SC  algorithm does not provide maximum likelihood decoding. A successive cancellation list (SCL) decoding algorithm was suggested in \cite{tal2015list},
and shown to achieve substantially better performance with complexity $O(Ln\log n)$.  large values of $L$ are needed  to implement near-ML\ decoding of polar subcodes and polar codes with CRC. This makes practical implementations of such list decoders very challenging. 

In practice one does not need to obtain a list of codewords, but just a single most probable one. The  Tal-Vardy algorithm for polar codes with CRC examines the elements in the obtained list, and discards those with invalid checksums. This algorithm can be easily tailored to process the dynamic freezing constraints used in the construction of polar subcodes \cite{trifonov2016polar} before the decoder reaches the last phase, so that the output list contains only valid codewords. However, even in this case  $L-1$ codewords are discarded from the obtained list, so most of the  work performed by the Tal-Vardy decoder is just wasted. 

This problem was addressed in \cite{niu2012stack}, where a generalization of the stack algorithm to the case of polar codes was suggested. It provides lower average decoding complexity compared to the Tal-Vardy algorithm. In this paper we revise the stack decoding algorithm for polar (sub)codes, and show that its complexity can be substantially reduced.  
\subsection{Stack decoding algorithm}
\label{sSeqDecAlg}
Let $u_0^{n-1}$ be the input vector used by the transmitter. Given a received noisy vector $y_0^{n-1}$, the proposed decoding algorithm constructs sequentially a number of partial candidate information vectors $v_0^{\phi-1}\in \F_2^\phi, \phi\leq n$, evaluates how close their continuations $v_0^{n-1}$ may be to the received sequence, and eventually produces a single codeword, being a solution of the decoding problem.


The stack  decoding algorithm  \cite{Zigangirov1966some,johannesson1998fundamentals,niu2012stack,miloslavskaya2014sequential}
employs a  priority queue\footnote{A PQ is commonly called "stack" in the sequential decoding literature. However, the implementation of the considered algorithm relies on Tal-Vardy data structures \cite{tal2015list}, which make use of the true stacks. Therefore, we employ the standard terminology of computer science.} (PQ) to store paths together with their scores.
A PQ is a data structure, which contains tuples $(M,v_0^{\phi-1})$, where $M=M(v_0^{\phi-1},y_0^{n-1})$ is the score of path $v_0^{\phi-1}$,  and provides efficient algorithms for the following operations \cite{Cormen2001introduction}:
\begin{itemize}
\item push a tuple into the PQ;
\item pop  a tuple $(M,v_0^{\phi-1})$  with the highest $M$;
\item remove  a given tuple from the PQ.
\end{itemize}
We assume here that the PQ may contain at most $D$ elements.

In the context of polar codes, the stack decoding algorithm operates as follows:
\begin{enumerate}
\item Push into the PQ the root of the tree with score $0$. Let $t_0^{n-1}=0$.
\item Extract from the PQ a path $v_0^{\phi-1}$ with the highest score. Let $t_\phi\gets t_\phi+1$.
\item If $\phi=n$, return codeword $v_0^{n-1}A_{m}$ and terminate.
\item If  the number of valid (i.e. those satisfying freezing constraints) children $v_0^{\phi}$ of path $v_0^{\phi-1}$ exceeds the amount of free space in the PQ, remove from it the element with the smallest score.
\item Compute the scores $M(v_0^\phi,y_0^{n-1})$ of  valid children $v_0^\phi$ of the extracted path, and push them into the PQ. 
\item If $t_{\phi}\geq L$, remove from PQ all paths $v_0^{j-1}, j\leq \phi$.
\item Go to step 2.
\end{enumerate}
In what follows, one iteration means one pass of the above algorithm over steps 2--7. Variables $t_\phi$ are used to ensure that the worst-case complexity of the algorithm does not exceed that of a list SC decoder with list size $L$.

The parameter $L$ has the same impact on the performance of the decoding algorithm as the list size in the Tal-Vardy  algorithm, since it imposes an upper bound on number of paths $t_\phi$ considered by the decoder at each phase $\phi$. Step 6 ensures that the algorithm terminates in at most $Ln$ iterations. This is also an upper bound on the number of entries stored in the PQ. However, the algorithm can work with PQ of much smaller size $D$. Step 4 ensures that this size is never exceeded.
\subsection{Score function}
There are many possible ways to define a score function for sequential decoding. In general, this should be done so that one can perform meaningful comparison of paths $v_0^{\phi-1}$ of different length $\phi$. The classical Fano metric for sequential decoding of convolutional codes is given by the probability $$P\set{\mathcal M|y_0^{n-1}}=\frac{P\set{\mathcal M,y_0^{n-1}}}{\prod_{i=0}^{n-1}W(y_i)},$$ where $\mathcal M$ is a variable-length message (i.e. a path in the code tree), and $W(y_i)$ is the probability measure induced on the channel output
alphabet when the channel inputs follow some prescribed (e.g. uniform) distribution \cite{massey1972variable}. In the context of polar codes, a straightforward implementation of this approach would correspond to score function $$M_1(v_0^{\phi-1},y_0^{n-1})=\log W_m^{(\phi-1)}\set{v_0^{\phi-1}|y_0^{n-1}}.$$
This is exactly the score function used in \cite{niu2012stack}.  However, there are several shortcomings in such definition:
\begin{enumerate}
\item Although the value of the score does depend on all $y_i, 0\leq i<n$, it does not take into account freezing constraints on symbols $u_i, i\in \mF,i\geq \phi$. As a result, there may exist  incorrect paths $v_0^{\phi-1}\neq u_0^{\phi-1}$, which have many low-probability continuations $v_0^{n-1},v_\phi^{n-1}\in \F_2^{n-\phi}$, such that the probability $$W_m^{(\phi-1)}\set{v_0^{\phi-1}|y_0^{n-1}}=\sum_{v_\phi^{n-1}}W_m^{(n-1)}\set{v_0^{n-1}|y_0^{n-1}}$$
becomes  high, and the stack decoder is forced to expand such a path. This is not a problem for convolutional codes, where the decoder may recover after an error burst, i.e. obtain a codeword identical to the transmitted one, except for a few closely located symbols. \item Due to freezing constraints, not all vectors $v_0^{\phi-1}$ correspond to valid paths in the code tree. This does not allow one to fairly compare the probabilities of paths of different lengths with  different number of frozen symbols. 
\item Computing probabilities $W_m^{(\phi-1)}\set{v_0^{\phi-1}|y_0^{n-1}}$ involves expensive multiplications and is prone to numeric errors. 
\end{enumerate}

The first of the above problems can be addressed by considering only the most probable continuation of path $v_0^{\phi-1}$, i.e. the score function can be defined as
\begin{equation*}
M_2(v_0^{\phi-1},y_0^{n-1})=\max_{v_\phi^{n-1}\in \F_2^{n-\phi}}\log W_m^{(n-1)}\set{v_0^{n-1}|y_0^{n-1}}.
\end{equation*}
Observe that maximization is performed over last $n-\phi$ elements of vector $v_0^{n-1}$, while the remaining ones are given by $v_0^{\phi-1}$.
 Let us further define $$\mathbf V(v_0^{\phi-1},y_0^{n-1})=\arg\max_{\substack{w_0^{n-1}\in \F_2^{n}\\w_0^{\phi-1}=v_0^{\phi-1}}}\log W_m^{(n-1)}\set{w_0^{n-1}|y_0^{n-1}},$$
i.e. $M_2(v_0^{\phi-1},y_0^{n-1})=\log W_m^{(n-1)}\set{\mathbf V(v_0^{\phi-1},y_0^{n-1})|y_0^{n-1}}.$
 
As  shown below, employing such score function already provides significant reduction of the average number of iterations at the expense of a negligible performance degradation. Furthermore, it turns out that this score  is exactly equal to the one used in the min-sum version of the Tal-Vardy list decoding algorithm \cite{balatsoukasstimming2015llrbased}, i.e. it can be computed in a very simple way. 

To address the second problem, we need to evaluate the probabilities of vectors $v_0^{n-1}$ under freezing conditions. To do this,  consider the set $C(\phi)$ of valid length-$\phi$ prefixes of the input vectors of the polarizing transformation, i.e. vectors $v_0^{\phi-1}$ satisfying the corresponding freezing constraints. 
Let us further define the set of their most likely continuations, i.e. $$\overline  C(\phi)=\set{\mathbf V(v_0^{\phi-1},y_0^{n-1})|v_0^{\phi-1}\in C(\phi)}.$$

For any $v_0^{n-1}\in \overline C(\phi)$ the probability of transmission of  $v_0^{n-1}A_m$, under condition  of $v_0^{\phi-1}\in  C(\phi)$ and   given the received vector $y_0^{n-1}$, equals 
\begin{align*}
{\mathbb W\set{v_0^{n-1}|y_0^{n-1},C(\phi)}=}
\frac{W_m^{(n-1)}\set{U_0^{n-1}=v_0^{n-1}| y_0^{n-1}}}{W_m^{(n-1)}\set{U_0^{n-1}\in \overline C(\phi)| y_0^{n-1}}}.
\end{align*}

Hence, an ideal score function could be defined as $$\mathbb M(v_0^{\phi-1},y_0^{n-1})=\log \mathbb W\set{\mathbf V(v_0^{\phi-1},y_0^{n-1})|y_0^{n-1},C(\phi)}.$$
Observe that this function is defined only for vectors $v_0^{\phi-1}\in C(\phi)$, i.e. those satisfying freezing constraints up to phase $\phi$.

Unfortunately, there is no simple and obvious way to compute $\pi(\phi,y_0^{n-1})=W_m^{(n-1)}\set{U_0^{n-1}\in \overline C(\phi)|y_0^{n-1}}$. Therefore, we have to develop an approximation. 
  It can be seen that 
\begin{align}
\label{mCodeProb}
\pi(\phi,y_0^{n-1})=&   W_m^{(n-1)}\set{\mathbf V(u_0^{\phi-1})| y_0^{n-1}}+\nonumber\\&\underbrace{\sum_{\substack{v_0^{\phi-1}\in C(\phi) \\v_0^{\phi-1}\neq{u_0^{\phi-1}}}}W_m^{(n-1)}\set{\mathbf V(v_0^{\phi-1})|y_0^{n-1}}}_{\mu(u_0^{\phi-1},y_0^{n-1})}.
\end{align}
Observe that $p=\E{\mathbf Y}{\frac{\mu(u_0^{n-1},\mathbf Y)}{\pi(\phi,\mathbf Y)}}$ is the total probability of incorrect paths at phase $\phi$ of the min-sum version of the Tal-Vardy list decoding algorithm with infinite list size, under the condition of all processed freezing constraints  being satisfied.  We consider decoding of polar (sub)codes, which are constructed to have low list SC decoding error probability even for small list size in the considered channel $W(y|c)$. Hence, it can be assumed that $p\ll 1$. This implies that with high probability $\mu(u_0^{\phi-1},y_0^{n-1})\ll W_m^{(n-1)}\set{\mathbf V(u_0^{\phi-1})| y_0^{n-1}}$, i.e. $\pi(\phi,y_0^{n-1})\approx    W_m^{(n-1)}\set{U_0^{n-1}=\mathbf V(u_0^{\phi-1})| y_0^{n-1}}$. 

   However, a real decoder cannot compute this value, since the transmitted vector $u_0^{n-1}$ is not available at the receiver side.  Therefore, we  propose to further approximate the logarithm of the first term in \eqref{mCodeProb} with its expected value over $\mathbf Y$, i.e.
\begin{equation}
\label{mScoreFunction}
\log \pi(\phi,y_0^{n-1})\approx \Psi(\phi)=\E{\mathbf Y}{\log W_m^{(\phi-1)}\set{\mathbf V(u_0^{\phi-1})|\mathbf Y}}
\end{equation}
Observe that this value depends only on $\phi$ and underlying channel $W(Y|C)$, and can be pre-computed offline.

Hence, instead of the ideal score function $\mathbb M(v_0^{\phi-1},y_0^{n-1})$, we propose to use an approximate one 
\begin{equation}
\label{mPathScore}
M_3(v_0^{\phi-1},y_0^{n-1})=M_2(v_0^{\phi-1},y_0^{n-1})-\Psi(\phi).
\end{equation}
Observe that for the correct path $v_0^{\phi-1}=u_0^{\phi-1}$ one has $\E{\mathbf Y}{M_3(v_0^{\phi-1},\mathbf Y)}=0$.
\subsection{Computing the score function}
Consider  computing  
\begin{align*}
R_m^{(\phi-1)}(v_0^{\phi-1},y_0^{n-1})=&M_2(v_0^{\phi-1},y_0^{n-1})\\=&\max_{\substack{w_0^{n-1}\in \F_2^{n}\\w_0^{\phi-1}=v_0^{\phi-1}}}\log W_m^{(n-1)}\set{w_0^{n-1}|y_0^{n-1}},
\end{align*}
  
Let  the modified log-likelihood ratios be defined as
\begin{equation}
\label{mLLRS}
S_m^{(\phi)}(v_0^{\phi-1},y_0^{n-1})=R_m^{(\phi)}(v_0^{\phi-1}.0,y_0^{n-1})-R_m^{(\phi)}(v_0^{\phi-1}.1,y_0^{n-1}).
\end{equation}
It can be seen that
\begin{align}
R_m^{(\phi)}(v_0^{\phi},y_0^{n-1})=&
R_m^{(\phi-1)}(v_0^{\phi-1},y_0^{n-1})\nonumber\\&+\tau(S_m^{(\phi)}(v_0^{\phi-1},y_0^{n-1}),v_\phi), \label{mLogProb}
\end{align}
where 
$$\tau(S,v)=\begin{cases}
0,&\text{if $\sgn(S)=(-1)^v$}\\
-|S|,&\text{otherwise.}
\end{cases}$$
is the penalty function.
Indeed, let $\tilde v_0^{n-1}=\mathbf V(v_0^{\phi-1})$. If $v_\phi=\tilde v_\phi$, then the most probable continuations of  $v_0^{\phi-1}$ and $v_0^{\phi}$ are identical.
Otherwise, $-\left|S_m^{(\phi)}(v_0^{\phi-1},y_0^{n-1})\right|$ is exactly the difference between the log-probability of the most likely continuations of $v_0^{\phi-1}$ and $v_0^{\phi}$. 

The initial value for recursion \eqref{mLogProb} is given by $$R_m^{(-1)}(y_0^{n-1})=\log \prod_{i=0}^{n-1}W\set{C=\hat c_i|Y=y_i},$$
where $\hat c_i$ is the hard decision  corresponding to $y_i$. However, this value can be replaced with $0$, since it does not affect the selection of paths in the stack algorithm. 

Efficient techniques for computing modified LLRs $S_m^{(\phi)}(v_0^{\phi-1},y_0^{n-1})$ for some kernels were derived in \cite{trofimiuk2019reduced}.

\subsection{Modified LLRs for Arikan kernel}
Since $R_m^{(\phi-1)}(v_0^{\phi-1},y_0^{n-1})$ is obtained by maximization of $W_m^{(n-1)}\set{ v_0^{n-1}|y_0^{n-1}}$ over $v_\phi^{n-1}$, it can be seen that 
\begin{align*} 
\lefteqn{R_{\lambda}^{(2i)}(v_0^{2i},y_0^{N-1})=}&\nonumber\\
&\max_{v_{2i+1}} \left(R_{\lambda-1}^{(i)}\left(v_{0,e}^{2i+1}\oplus v_{0,o}^{2i+1},y_0^{\frac{N}{2}-1}\right)+R_{\lambda-1}^{(i)}\left(v_{0,o}^{2i+1},y_{\frac{N}{2}}^{N-1}\right)\right),\\
\lefteqn{R_{\lambda}^{(2i+1)}(v_0^{2i+1}|y_0^{N-1})=}&\nonumber\\
&\quad \quad \quad R_{\lambda-1}^{(i)}\left(v_{0,e}^{2i+1}\oplus v_{0,o}^{2i+1},y_0^{\frac{N}{2}-1}\right)+R_{\lambda-1}^{(i)}\left(v_{0,o}^{2i+1},y_{\frac{N}{2}}^{N-1}\right),
\end{align*}
where $N=2^\lambda,$ $0< \lambda\leq m$, and initial values for these recursive expressions are given by $R_0^{(0)}(b,y_j)=\log W_0^{(0)}\set{b|y_j}$, $b\in\set{0,1}$. 
From \eqref{mLLRS} one obtains 
\begin{align*}
S_{\lambda}^{(2i)}(v_0^{2i-1}|y_0^{2^{\lambda}-1})=&\max(J(0)+K(0),J(1)+K(1))-\nonumber\\&\max(J(1)+K(0),J(0)+K(1))\nonumber\\=
&\max(J(0)-J(1)+K(0)-K(1),0)-\nonumber\\&\max(K(0)-K(1),J(0)-J(1))\\
S_{\lambda}^{(2i+1)}(v_0^{2i},y_0^{2^{\lambda}-1})=&J(v_{2i})+K(0)-J(v_{2i}+1)-K(1)
\end{align*}
where $J(c)=R_{\lambda-1}^{(i)}((v_{0,e}^{2i-1}\oplus v_{0,o}^{2i-1}).c|y_0^{2^{\lambda-1}-1})$, $K(c)=R_{\lambda-1}^{(i)}(v_{0,o}^{2i-1}.c|y_{2^{\lambda-1}}^{2^{\lambda}-1})$. Observe that $$J(0)-J(1)=a=S_{\lambda-1}^{(i)}(v_{0,e}^{2i-1}\oplus v_{0,o}^{2i-1},y_0^{2^{\lambda-1}-1})$$ and $$K(0)-K(1)=b=S_{\lambda-1}^{(i)}(v_{0,o}^{2i-1},y_{2^{\lambda-1}}^{2^\lambda-1})$$

It can be obtained from these expressions that the modified log-likelihood ratios are given by 
\begin{align}
\label{mMinSum1}
S_{\lambda}^{(2i)}(v_0^{2i-1},y_0^{2^\lambda-1})=&Q(a,b)=\sgn (a)\sgn (b)\min(|a|,|b|),\\
\label{mMinSum2}
S_{\lambda}^{(2i+1)}(v_0^{2i},y_0^{2^\lambda-1})=&P(v_{2i},a,b)=(-1)^{v_{2i}}a+b.
\end{align}
The initial values for this recursion are given by $S_0^{(0)}(y_i)=\log\frac{W\set{0|y_i}}{W\set{1|y_i}}$. 
These expressions can be readily recognized as the min-sum approximation of the list SC algorithm\cite{balatsoukasstimming2015llrbased}. However, these
are also the exact values, which reflect the probability of the most likely continuation of a given path $v_0^{\phi-1}$ in the code tree. 

\subsection{The bias function}
The function  $\Psi(\phi)$ is equal to the expected value of the logarithm of the probability of a length-$\phi$ part of the correct path, i.e. the path corresponding to the vector $u_0^{n-1}$ used by the encoder. Employing this function enables one to estimate how far a particular path $v_0^{\phi-1}$ has diverted from the expected behaviour of a correct path. The bias function can be computed offline under the assumption of zero codeword transmission. 

In the  case of Arikan kernels the cumulative  density functions $F_{\lambda}^{(i)}(x)$ of $S_\lambda^{(i)}$ are given by \cite{kern2014new}
\begin{align*}
&F_{\lambda}^{(2i)}(x)=\begin{cases}
2F_{\lambda-1}^{(i)}(x)(1-F_{\lambda-1}^{(i)}(-x)),&x< 0\\
2F_{\lambda-1}^{(i)}(x)-(F_{\lambda-1}^{(i)}(-x))^2-(F_{\lambda-1}^{(i)}(x))^2,&x\geq 0
\end{cases}\\
&F_{\lambda}^{(2i+1)}(x)=\int_{-\infty}^\infty F_{\lambda-1}^{(i)}(x-y) dF_{\lambda-1}^{(i)}(y),
\end{align*}
where $F_0^{(0)}(x)$ is the CDF of the channel output LLRs.
Hence,
\begin{equation}
\label{mBias}
\Psi(\phi)=-\sum_{i=0}^{\phi-1} \int_{-\infty}^0F_{m}^{(i)}(x)dx.
\end{equation}

No simlpe expressions are available for computing the CDF of LLRs in the case of large kernels. Hence, the score function for such kernels can be computed by simulating genie-aided SC decoder, and averaging the score $M_2(u_0^{\phi-1},y_0^{n-1})$ for the correct path $u_0^{n-1}$.

The bias function $\Psi(\phi)$ depends only on $m$ and channel properties, so it can be used for decoding of any polar (sub)code of a given length.

\subsection{Complexity analysis}
The algorithm presented in Section \ref{sSeqDecAlg} extracts from the PQ length-$\phi$ paths at most $L$ times. At each iteration it needs to calculate the LLR $S_m^{(\phi)}(v_0^{\phi-1},y_0^{n-1})$. Intermediate values for these calculations can be reused in the same way as in \cite{tal2015list}. Hence, LLR calculations require at most $O(Ln\log n)$ operations. However, simulation results presented below suggest that the average complexity of the proposed algorithm is substantially lower, and at high SNR approaches $O(n\log n)$, the complexity of the SC algorithm.  

\section{Implementation}
In this section we present implementation details of the sequential decoding algorithm for the case of polar codes with Arikan kernel. We follow \cite{miloslavskaya2014sequential,trofimiuk2020fast}.

\begin{figure}
\small
\begin{algorithm}{Decode}{y_0^{n-1}, L, D}
\CALL{InitializeDataStructures}(D)\\
l\=\CALL{AssignInitialPath}()\\
\tilde S\=\CALL{GetArrayPointerS\_W}(l,0,0)\\
\tilde S[0..n-1]\=y_0^{n-1}\\
R_l\=0\\
\CALL{Push}(l,0)\\
\begin{WHILE}{true}
(M,l) \leftarrow \CALL{PopMax}()\\
\begin{IF}{\phi_l = n}
\RETURN{\CALL{GetArrayPointerC\_R}(l,0,0)}     
\end{IF}\\
q_{\phi_l}\=q_{\phi_l}+1\\
\CALL{IterativelyCalcS}(l,m,\phi_l)\\
\begin{IF}{\phi_l\in \mF}
\CALL{ContinuePathFrozen}(l)
\ELSE
\begin{IF}{P\geq D-1}
\text{Let $l_0$ be the path with the smallest score $M_0$}\\
\CALL{KillPath}(l_0)\\
\text{Remove $(M_0,l_0)$ from the PQ}
\end{IF}\\
\CALL{ContinuePathUnfrozen}(l)
\end{IF}\\
\begin{IF}{q_{\phi_l-1} \ge L}
\begin{FOR}{\text{All paths $l'$ stored in the PQ}}
\begin{IF}{\phi_l'\leq \phi_{l}}
\CALL{KillPath}(l')\\
\text{Remove $l'$ from the PQ}
\end{IF}
\end{FOR}
\end{IF}
\end{WHILE}
\end{algorithm}
\caption{The sequential decoding algorithm}
\label{fSeqAlg}
\end{figure}

\begin{figure}
\begin{subfigure}{0.5\textwidth}
\begin{algorithm}{ContinuePathFrozen}{l}
v\=\CALL{EvaluateDynFrozen}(l,\phi_l)\\
S\=\CALL{GetArrayPointerS\_R}(l,m)\\
C\=\CALL{GetArrayPointerC\_W}(l,m)\\
C[0]\=v\\
\CALL{UpdateDynFrozen}(l,v)\\
\begin{IF}{(-1)^vS[0]<0}
R_l\=R_l-|S[0]|
\end{IF}\\
\begin{IF}{\phi_l\equiv 1\bmod 2}
\CALL{IterativelyUpdateC}(l,m,\phi_l)
\end{IF}\\
\phi_l\=\phi_l+1\\
\CALL{Push}(l,R_l-\Psi(\phi_l))
\end{algorithm}
\caption{Frozen symbols}
\end{subfigure}
\begin{subfigure}{0.5\textwidth}
\begin{algorithm}{ContinuePathUnfrozen}{l}
S\=\CALL{GetArrayPointerS\_R}(l,m)\\
C\=\CALL{GetArrayPointerC\_W}(l,m)\\
C[0]\=S<0\\
\CALL{UpdateDynFrozen}(l,C[0])\\
l'\=\CALL{ClonePath}(l)\\
C'\=\CALL{GetArrayPointerC\_W}(l',m)\\
C'[0]=1-C[0]\\
\CALL{UpdateDynFrozen}(l',C'[0])\\
R_{l'}\=R_l-|S[0]|\\
\begin{IF}{\phi_l\equiv 1\bmod 2}
\CALL{IterativelyUpdateC}(l,m,\phi_l)\\
\CALL{IterativelyUpdateC}(l',m,\phi_l)
\end{IF}\\
\phi_l\=\phi_l+1; \phi_{l'}\=\phi_l\\
\CALL{Push}(l,R_l-\Psi(\phi_l))\\
\CALL{Push}(l',R_{l'}-\Psi(\phi_{l'}))
\end{algorithm}
\caption{Unfrozen symbols}
\end{subfigure}
\caption{Path update procedures}
\end{figure}

\begin{figure}
\small
\begin{subfigure}{0.5\textwidth}
\begin{algorithm}{IterativelyCalcS}{l,\lambda,\phi}
d\=\max\set{0\leq d'\leq \lambda-1|\phi \text{is divisible by $2^{d'}$}}\\
\lambda'\=\lambda-d\\
S'\=\CALL{GetArrayPointerS\_R}(l,\lambda'-1)\\
N\=2^{m-\lambda'}\\
\begin{IF}{\text{$\phi 2^{-d}$ is odd}}
\tilde C\=\CALL{GetArrayPointerC\_R}(l,\lambda')\\
S''\=\CALL{GetArrayPointerS\_W}(l,\lambda')\\
S''[\beta]\=\CALL{P}(\tilde C[\beta],S'[\beta],S'[\beta+N]), 0\leq \beta<N\\
S'\=S'';\lambda'\=\lambda'+1;N \leftarrow N/2
\end{IF}\\
\begin{WHILE}{\lambda'\leq \lambda}
S''\=\CALL{GetArrayPointerS\_W}(l,\lambda')\\
S''[\beta]\=\CALL{Q}(S'[\beta+N],S'[\beta]), 0\leq \beta<N\\
S'\=S'';\lambda'\=\lambda'+1;N\=N/2
\end{WHILE}
\end{algorithm}
\caption{Computing  $S_{\lambda}^{(\phi)}(v_0^{\phi-1},y_0^{N-1})$}
\label{fIterativelyCalcS}
\vspace{0.5cm}
\end{subfigure}
\small
\begin{subfigure}{0.5\textwidth}
\begin{algorithm}{IterativelyUpdateC}{l,\lambda,\phi}
\delta\=\max\set{d|\phi+1 \text{is divisible by
$2^{d}$}}\\
\tilde C\=\CALL{GetArrayPointerC\_W}(l,\lambda-\delta,0)\\
N\=2^{m-\lambda};
\tilde C=\tilde C+N(2^\delta-2); C''\=\tilde C+N;\\
\lambda'\=\lambda-\delta\\
\begin{WHILE}{\lambda>\lambda'}
C'\=\CALL{GetArrayPointerC\_R}(l,\lambda)\\
\tilde C[\beta]\=C'[\beta]\oplus C''[\beta], 0\leq \beta<N\\
N\=2N;
C''\=\tilde C;
\tilde C\=\tilde C-N\\
\lambda\=\lambda-1
\end{WHILE}
\end{algorithm}
\caption{Updating $C$ arrays}
\label{fIterativelyUpdateC}
\end{subfigure}
\caption{Computing LLRs and partial sums}
\end{figure}

Figure \ref{fSeqAlg} presents the sequential decoding algorithm. Its input values are the received noisy values $y_0^{n-1}$, maximal number of visits per phase $L$, and priority queue size $D$.
It makes use of modified Tal-Vardy data structures \cite{tal2015list}, which are discussed below.
Here $l$ denotes path index, $\phi_l$ is the phase (i.e. length) of the $l$-th path, $q_j$ is the number of times the decoder has extracted a path of length $j$, and $R_l$ is the accumulated penalty of the $l$-th path, given by \eqref{mLogProb}.

After initialization of the data structures, the received LLRs are loaded into the $0$-th layer of Tal-Vardy data structures\footnote{Technically, one needs to compute LLRs $S_0^{(0)}(y_i)=\ln\frac{W(0|y_i)}{W(1|y_i)}, 0\leq i<n$. However, one can scale appropriately the bias function, and avoid this calculation.}, and the initial path is pushed into the priority queue. At each iteration of the algorithm the index $l$ of a path $v_0^{\phi_l-1}$ with the maximal score $M$  is obtained. If its length is equal $n$, decoding terminates and the corresponding codeword is extracted at line 10.
Otherwise, phase visit counter is updated at line 11, and $S_m^{(\phi_l)}(v_0^{\phi_l-1},y_0^{n-1})$
is computed on line 12. If $\phi_l$ corresponds to a frozen symbol, then the path is appropriately extended on line 14. Otherwise, we need to ensure that there is sufficient amount of free space in the priority queue. If this is not the case, the worst path is removed at lines 17--18.  The path is cloned to obtain $v_0^{\phi_l-1}.0$ and $v_0^{\phi_l-1}.1$ at line 19. If the decoder exceeds the allowed number of visits per phase $L$, then all paths shorter than $\phi_l$ are eliminated at lines 23--24.

Procedures for extending the paths are shown in Figure \ref{fIterativelyUpdateC}. These procedures obtain pointer $S$ to the computed LLR $S_m^{(\phi_l)}(v_0^{\phi_l-1},y_0^{n-1})$, pointer to the extension symbol $C$ (and $C'$), and put there appropriate values.  For odd phases, the partial sum values are propagated to the lower layers via calls to $IterativelyUpdateC$.
The functions update the accumulated penalty values, and push the paths into the priority queue with appropriate score values. 

Here a call to $EvaluateDynFrozen$ corresponds to evaluation of \eqref{mDynFrozen}.
Some provisions are done for efficient evaluation of dynamic frozen symbols by calling $UpdateDynFrozen$.  Since for most practical codes the number of non-trivial dynamic freezing constraints (i.e. rows of matrix $V$    of weight more than 1) is small, in a software implementation one can store for each path a bitmask, so that each bit in it corresponds to a dynamic frozen symbol. The values of the bits are updated by $UpdateDynFrozen$ on the phases corresponding to non-zero values in the $V$ matrix. 

 Figures \ref{fIterativelyCalcS} and \ref{fIterativelyUpdateC} present
 iterative algorithms for computing $S_{l,\lambda}[\beta]$ and $C_{l,\lambda}[\beta]$.
 These algorithms resemble the recursive ones given in \cite{tal2015list}. However, the proposed implementation avoids costly array dereferencing operations. 
Each path is associated with arrays of intermediate LLRs $S_{l,\lambda}[\beta], 0\leq l<D, 0\leq \lambda\leq m,0\leq \beta<2^{m-\lambda},$ where $D$ is the maximal number of paths considered by the decoder (i.e. the maximal size of the PQ). 

It was suggested in \cite{tal2015list} to store the arrays of partial sum tuples $C_{l,\lambda}[\beta][\phi\bmod 2]$. We propose to rename these arrays to $C_{l,\lambda,\phi\bmod 2}[\beta]$. By examining the \textit{RecursivelyUpdateC} algorithm presented in \cite{tal2015list}, one can see that $C_{l,\lambda,1}[\beta]$ is just copied to $C_{l,\lambda-1,\psi}[2\beta+1]$ for some $\psi\in\set{0,1}$, and this copy operation terminates on some layer $\lambda'$. Observe that  $\lambda-\lambda'$
is equal to the maximal integer  $d$, such that $\phi+1$ is divisible by $2^d$.
 Therefore, we propose to co-locate $C_{l,\lambda,1}[\beta]$  with $C_{l,\lambda_0,0}[\beta]$. In this case the corresponding pointers are given by $C_{l,\lambda,1}=C_{l,\lambda_0,0}+2^{m-\lambda}(2^{\lambda-\lambda'}-1)$. This not only results in the reduction of the amount of data stored by a factor of two, but also enables one to   avoid "copy on write"  operation (see line 6 of Algorithm 9 in \cite{tal2015list}). Therefore, we write $C_{l,\lambda_0}$ instead of $C_{l,\lambda_0,0}$ in what follows.

We use the array pointer mechanism suggested in \cite{tal2015list}  to avoid data copying. However, we distinguish the case of read and write data access.  Retrieving  read-only  pointers is performed by functions \textit{GetArrayPointerC\_R}$(l,\lambda)$ and \textit{GetArrayPointerS\_R}$(l,\lambda)$ shown in Figure \ref{fROAccess}. 
Retrieving writable pointers is performed by function \textit{GetArrayPointerW}$(T,l,\lambda)$, where $T\in \set{'C','S'}$ shown in Figure \ref{fWAccess}. This function implements reference counting mechanism similar to that proposed in \cite{tal2015list}.

Many paths considered by the proposed algorithm share common values of $S_{l,\lambda}[\beta]$ and $C_{l,\lambda}[\beta]$, similarly to SCL decoding. To avoid duplicate calculations
one can use the same shared memory data structures. That is, for each path $l$ and for each layer $\lambda$ we store the index of the array containing the corresponding values $S_{l,\lambda}[\beta]$ and $C_{l,\lambda}[\beta]$. This index is given by $p=$\textit{PathIndex2ArrayIndex}$[l,\lambda]$, so that the corresponding data can be accessed as \textit{ArrayPointer}$[T][p], T\in\set{'S','C'}$.
Furthermore, for each integer $p$ we maintain the number of references to this array \textit{ArrayReferenceCount}$[p]$. If the decoder needs to write the data into an array, which is referenced by more than one path, a new array needs to be allocated. Observe that there is no need to copy anything into this array, since it will be immediately overwritten. This is an important advantage with respect to the implementation described in \cite{tal2015list}. However, the sequence of array read/write and stack push/pop  operations still satisfies the validity assumptions introduced in \cite{tal2015list}, so the proposed algorithm can be shown to be well-defined by exactly the same reasoning as the original SCL.

\begin{figure}
\small
\parbox{0.5\textwidth}{
\begin{algorithm}{GetArrayPointerW}{T,l,\lambda}
p\=PathIndex2ArrayIndex[l,\lambda]\\
\begin{IF}{p=-1}
p\=\CALL{Allocate}(\lambda)
\ELSE
\begin{IF}{ArrayReferenceCount[p]>1}
ArrayReferenceCount[p]--\\
p\=\CALL{Allocate}(\lambda)
\end{IF} 
\end{IF}\\
\RETURN ArrayPointer[T][p]\ 
\end{algorithm}}
{
\begin{algorithm}{GetArrayPointerS\_W}{l,\lambda}
\RETURN \CALL{GetArrayPointerW}('S',l,\lambda)
\end{algorithm}
\begin{algorithm}{GetArrayPointerC\_W}{l,\lambda,\phi}
\delta\=\max\set{d|\phi+1 \text{is divisible by
$2^{d}$}}\\
C\=\CALL{GetArrayPointerW}('C',l,\lambda-\delta)\\
\begin{IF}{\phi\equiv 1\bmod 2}
C\=C+2^{m-\lambda}(2^{\delta}-1)
\end{IF}\\
\RETURN C
\end{algorithm}}
\caption{Write access to the data}
\label{fWAccess}
\end{figure}

\begin{figure}
\small
\begin{algorithm}{GetArrayPointerS\_R}{l,\lambda}
\RETURN\ ArrayPointer['S'][PathIndex2ArrayIndex[l,\lambda]]
\end{algorithm}
{
\begin{algorithm}{GetArrayPointerC\_R}{l,\lambda}
\RETURN\ ArrayPointer['C'][PathIndex2ArrayIndex[l,\lambda]]
\end{algorithm}
}
\caption{Read-only access to the data}
\label{fROAccess}
\end{figure}

The procedures $ClonePath$ and $KillPath$ operate in exactly the same way as in \cite{tal2015list}, except that they additionally keep track of the number $P$ of active paths.

The Tal-Vardy data structures need to be initialized to accomodate up to $D$ paths. This is implemented in $InitializeDataStructures$. This function additionally resets counters $q_0^{n-1}$.

An efficient implementation of the priority queue is given in \cite{yakuba2015multilevel}.

\section{Simulation results}
\begin{figure*}
\includegraphics[width=0.9\textwidth]{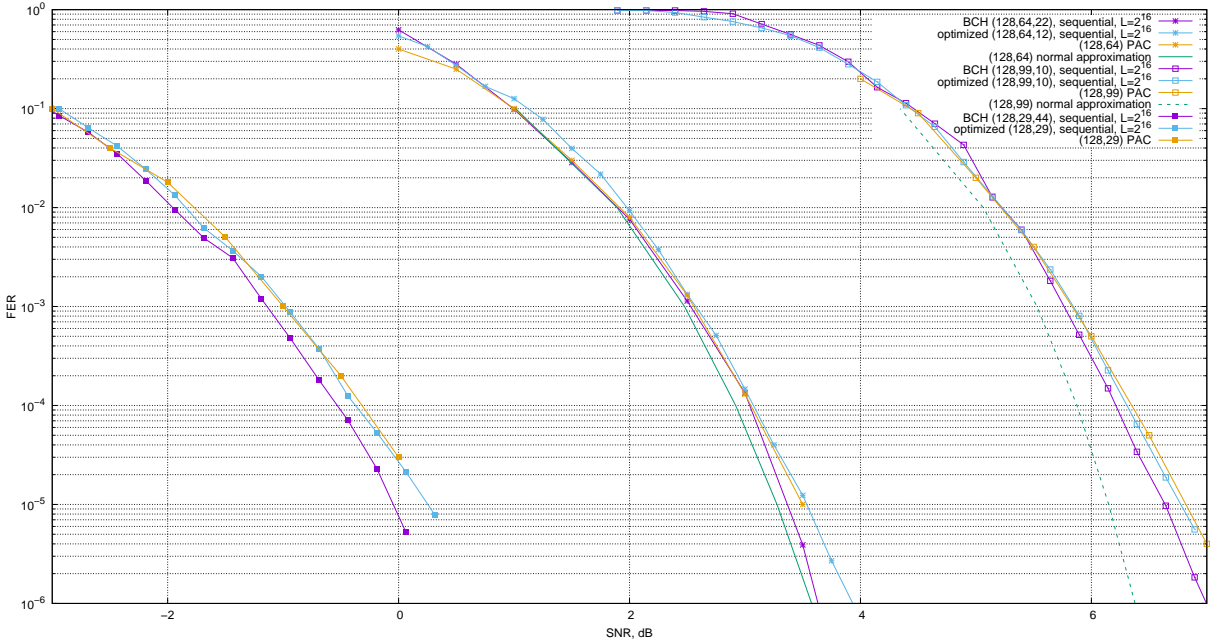}
\caption{Performance of codes of length 128}
\label{f128FER}
\end{figure*}

Figure \ref{f128FER} presents the performance of some codes of length $64$. We consider optimized polar subcodes \cite{trifonov2020randomized}, extended BCH\ codes, represented via \eqref{mCheckConstraint}, and Arikan PAC\ codes, reproduced from \cite{moradi2020performance}.  For comparison, we present also normal approximation\footnote{For $k=29$ the normal approximation appears to be rather imprecise.} for Polyanskiy-Poor-Verdu lower bound \cite{polyanskiy2010channel}
List size for the sequential decoder was set to $L=2^{16}$, so that almost no paths are killed.  
\begin{figure*}[t]
\begin{subfigure}{\textwidth}
\includegraphics[width=0.9\textwidth]{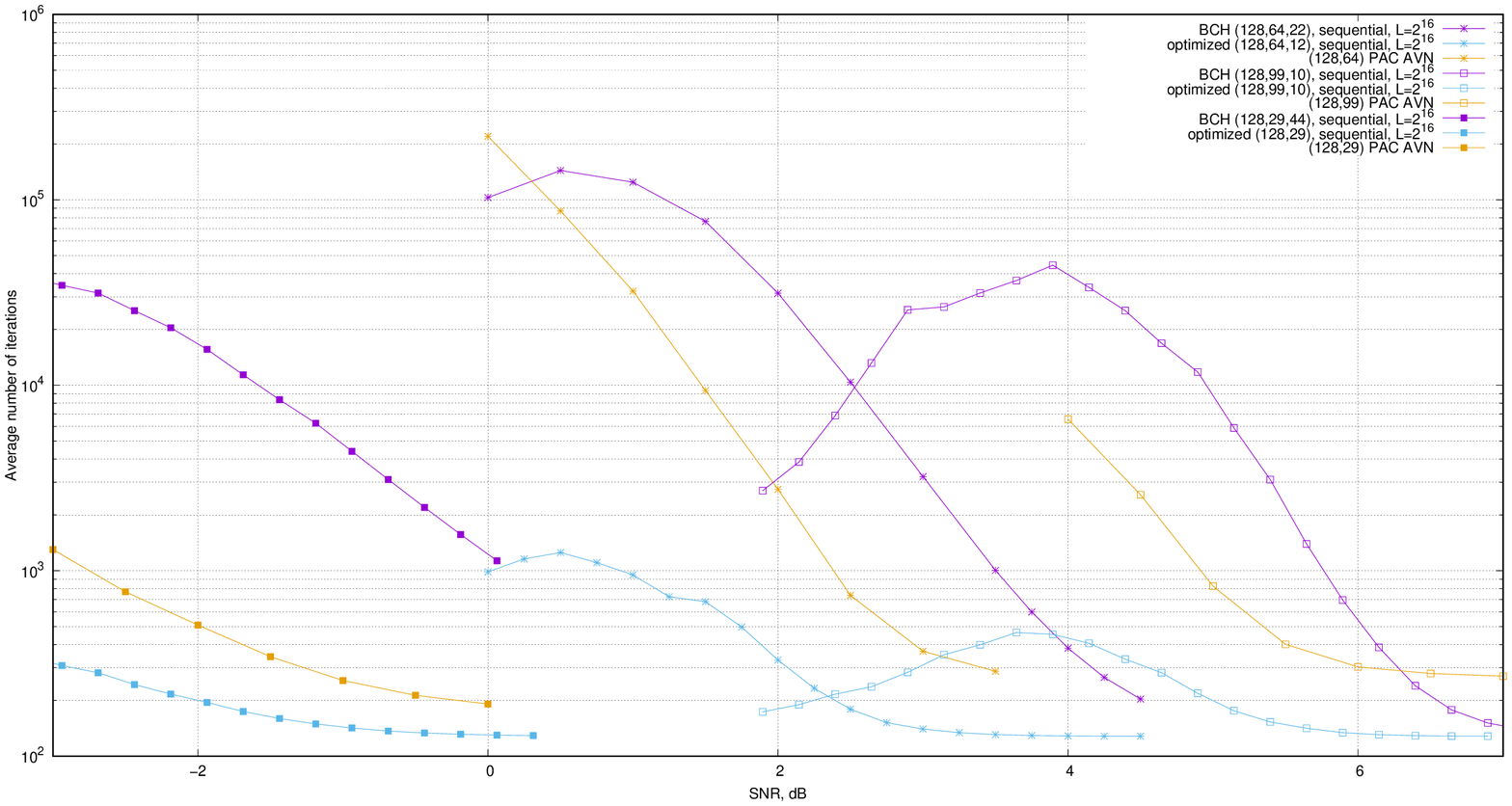}
\caption{Average number of sequential decoder iterations }
\label{f128AvgIt}
\end{subfigure}
\begin{subfigure}{\textwidth}
\includegraphics[width=0.9\textwidth]{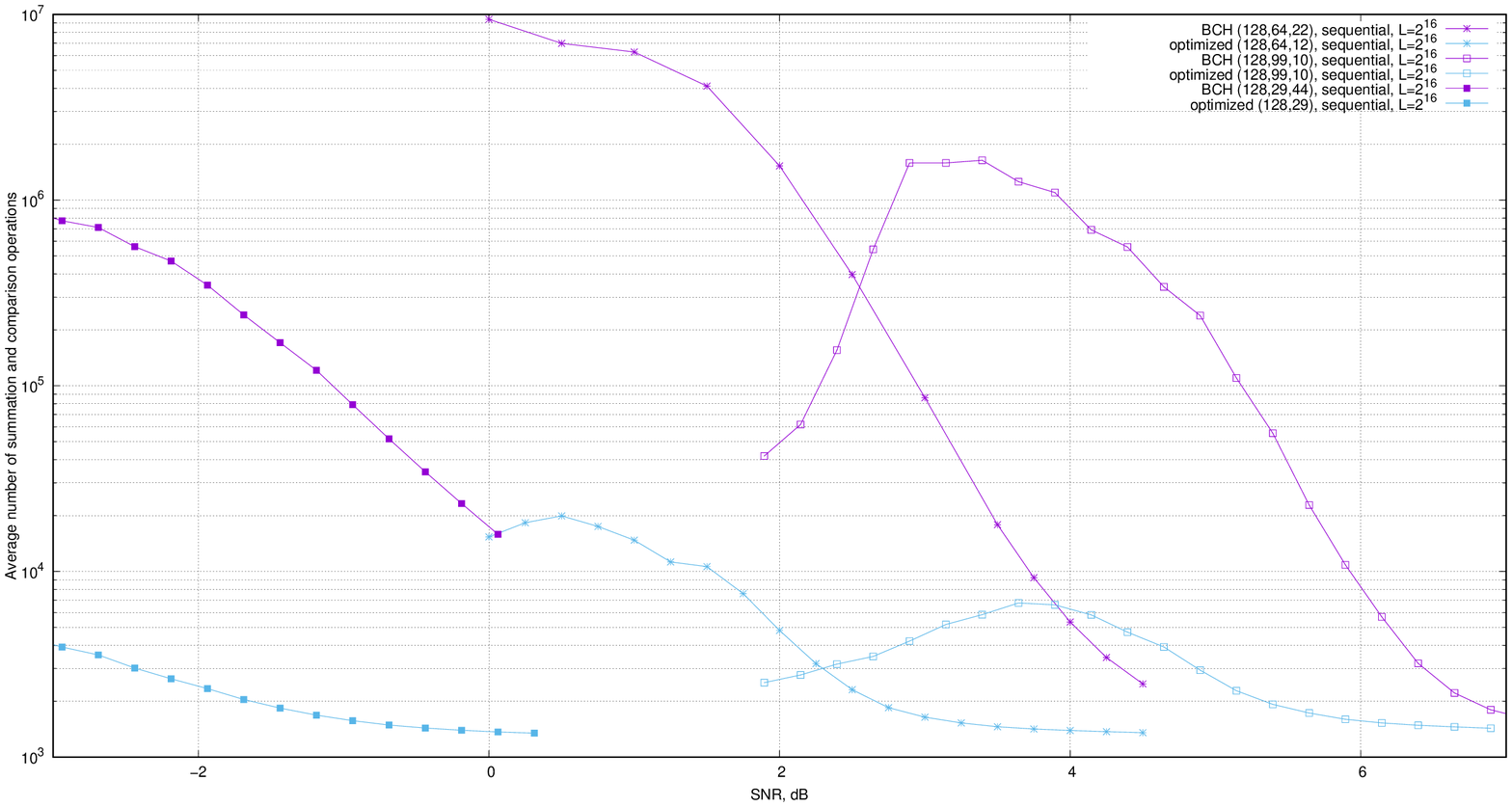}
\caption{Average number of arithmetic operations }
\label{f128AvgComp}
\end{subfigure}
\caption{Decoding complexity for codes of length 128}
\end{figure*}
Figure \ref{f128AvgIt} presents the associated decoding complexity.

It can be seen that extended BCH\ codes provide better performance compared to PAC codes. However, their average decoding comlpexity is rather high.  Optimized polar subcodes provide performance extremely close to that of PAC codes, but have much lower average decoding complexity. Observe that in the latter case the average number of iterations performed by the sequential decoder is much less compared to the average number of nodes visited by the decoder considered in  \cite{moradi2020performance}.
Figure \ref{f128AvgComp} presents the same results in terms of the number of arithmetic operations. Observe that the sequential decoding algorithm uses only summation and comparison operations. It can be seen that their average number converges to $n\log_2 n+C$, where $C$ is a small value corresponding to the cost of priority queue operations.

\begin{figure}
\includegraphics[width=0.5\textwidth]{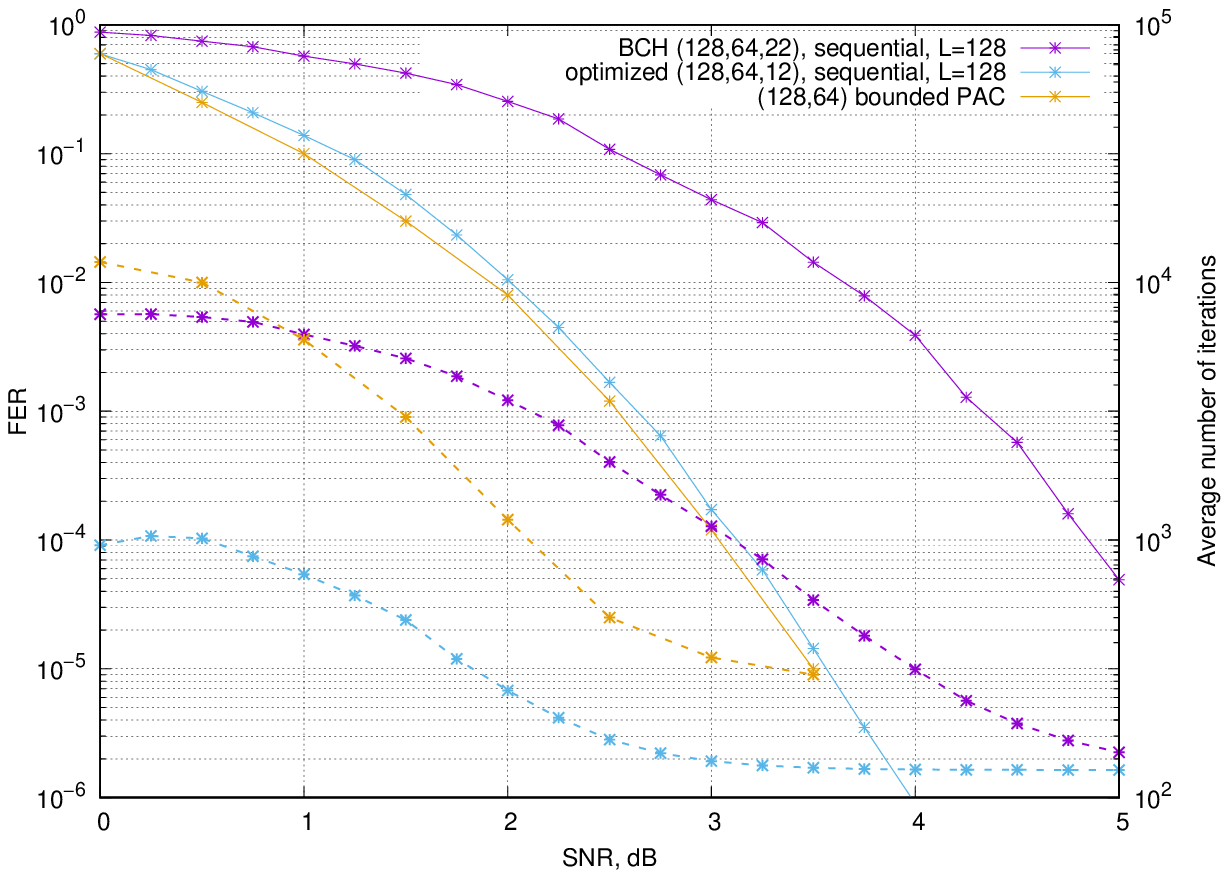}
\caption{Performance and decoding complexity of $(128,64)$ codes}
\label{f128FERLimited}
\end{figure}

Figure \ref{f128FERLimited} presents the performance and complexity of sequential decoder with $L=128$ and PAC\ decoder with maximal number of node visits set to $16384$ \cite{moradi2020performance}. It can be seen that the sequential decoder has much lower average complexity. Observe also that at high SNR the average number of iterations performed by the sequential decoder converges to $n$, the length of the code, while the average number of node visits for PAC decoder converges to a higher value.

\begin{figure*}
\includegraphics[width=0.5\textwidth]{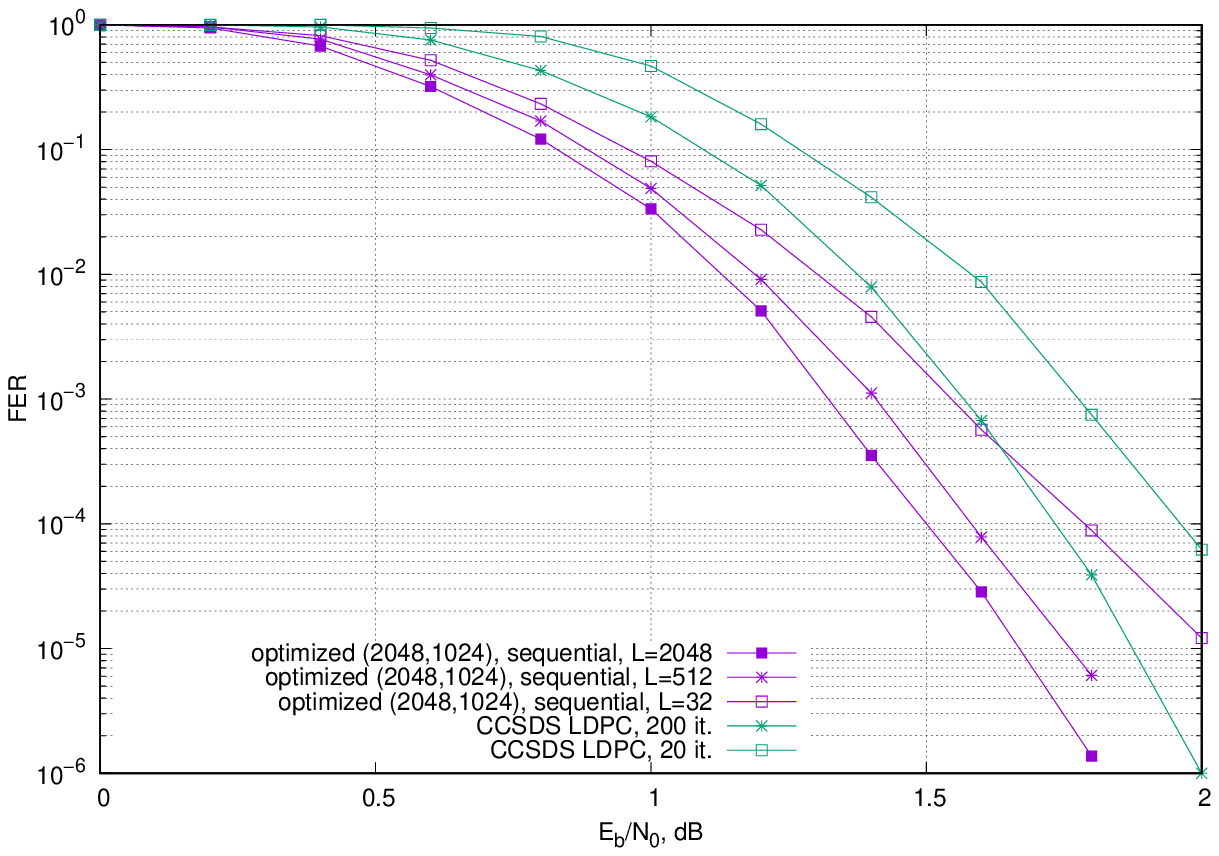}
\includegraphics[width=0.5\textwidth]{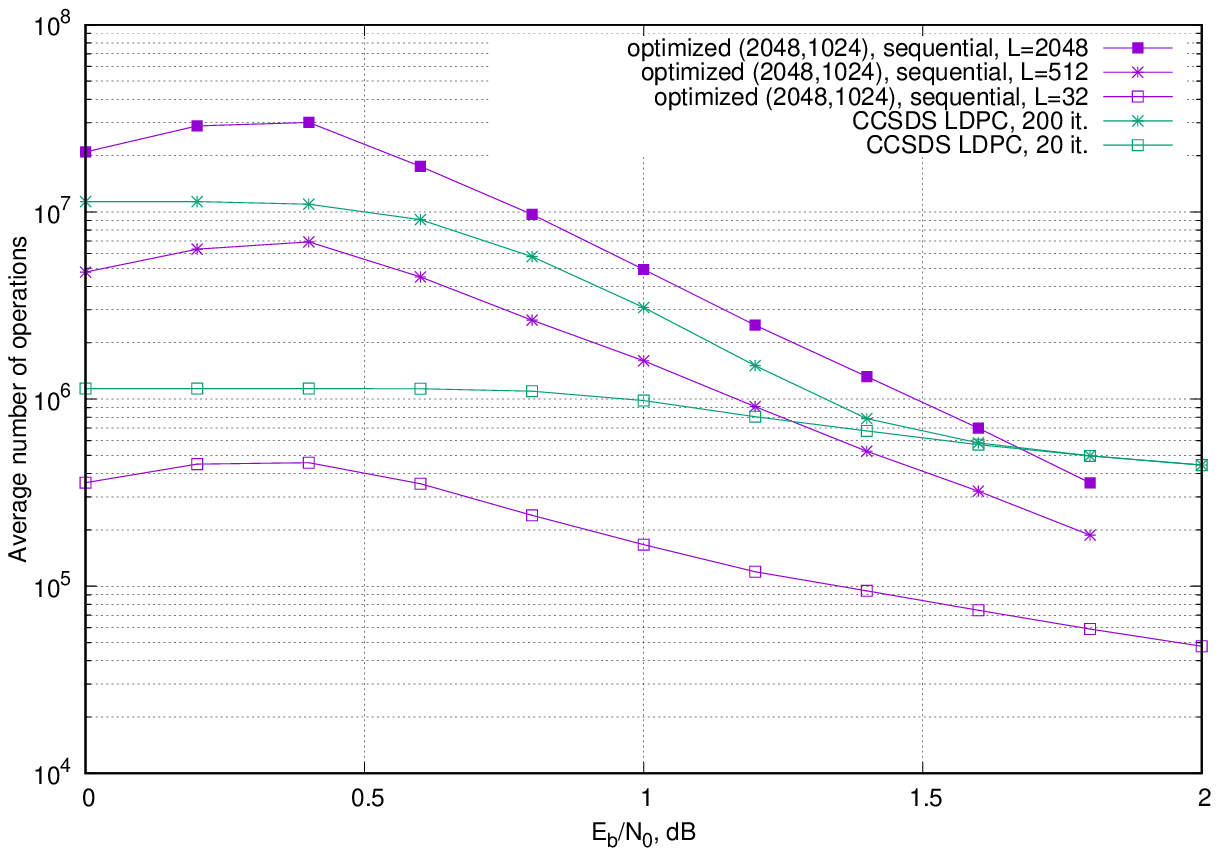}
\caption{Performance and decoding complexity of $(2048,1024)$ codes}
\label{f2048}
\end{figure*}

Figure \ref{f2048} presents performance and average decoding complexity of a $(2048,1024)$ optimized polar subcode, as well as CCSDS LDPC code under shuffled BP decoding \cite{zhang2005shuffled}. It can be seen that even with $L=32$ the sequential decoding algorithm provides performance close to that of shuffled BP decoder with 200 iterations. Further 0.15 dB gain is obtained by setting $L=512$.  Even in the latter case the sequential decoder requires in average less operations than the shuffled BP\ decoder for $E_b/N_0>1.3$ dB. Observe also that shuffled BP\ decoder requires computing $\log\tanh(.)$ function\footnote{Replacing it with some approximations may result in performance degradation}, while the sequential decoder uses only summation and comparison operations. Hence, the considered polar subcode under sequential decoding provides both better performance AND lower decoding complexity compared to the LDPC\ code.

Observe also, that increasing the number of iterations in the shuffled BP\ decoder does not provide any noticeable improvement. However, it is possible  to obtain further 0.1 dB gain with polar subcode by setting $L=2048$ in the sequential decoder. In this case the complexity of the sequential decoder becomes lower than the complexity of the shuffled BP\ decoder for $E_b/N_0>1.7$ dB. Even for $L=2048$ no maximum likelihood decoding errors were observed in sequential decoder simulations, so further performance improvement can be obtained by increasing decoding complexity.

\begin{figure*}
\includegraphics[width=0.5\textwidth]{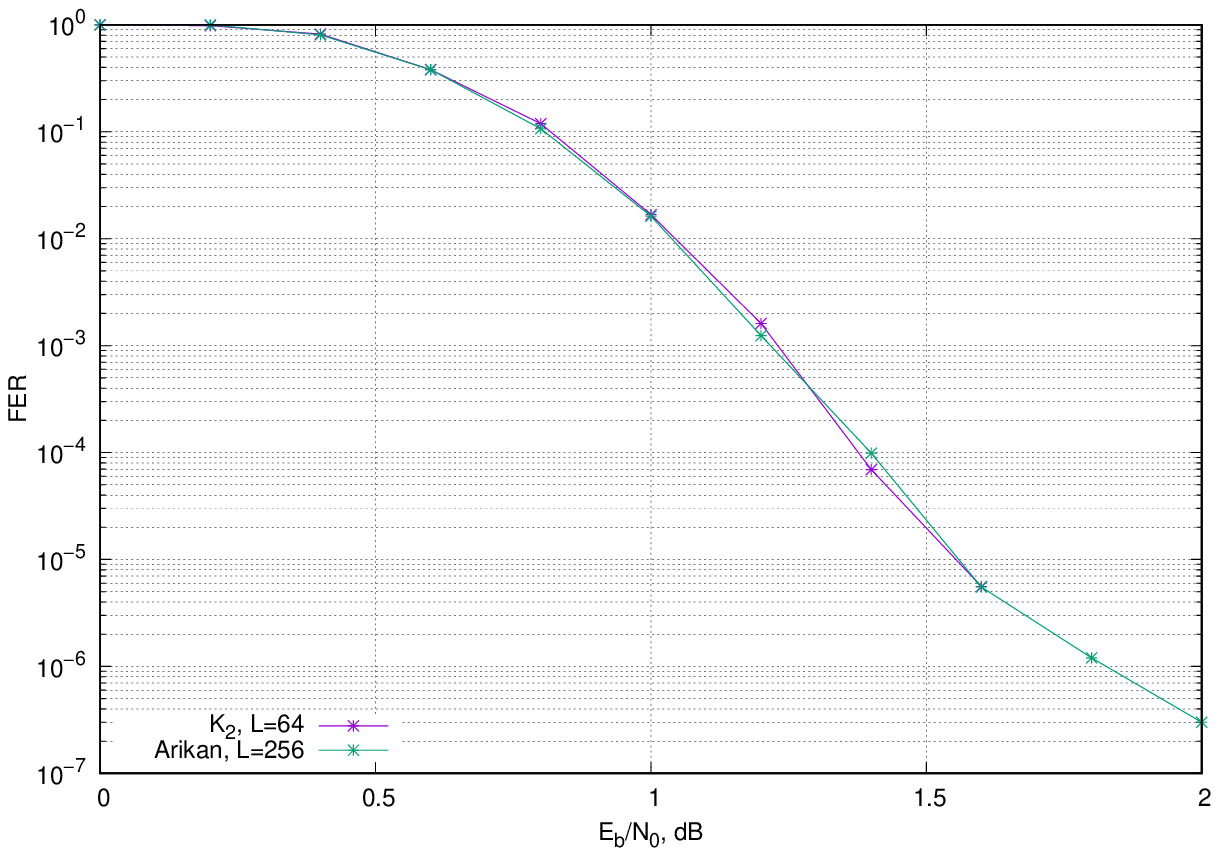}
\includegraphics[width=0.5\textwidth]{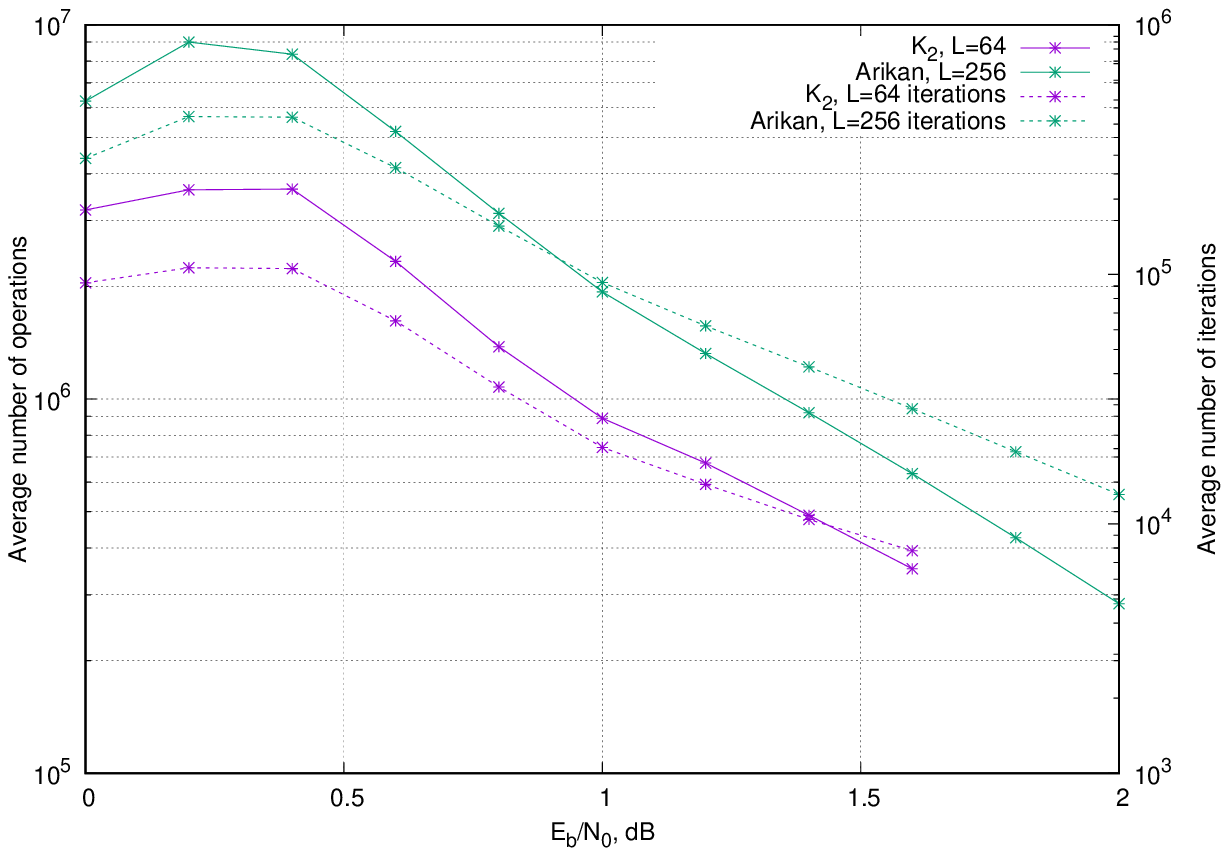}
\caption{Performance and decoding complexity of $(4096,2048)$ codes}
\label{f4096}
\end{figure*}
Figure \ref{f4096} presents performance and decoding complexity of randomized $(4096,2048)$ polar subcodes with Arikan kernel and $16\times 16$ kernel $K_2$ reported in \cite{trofimiuk2019reduced}. It can be seen that the code with large kernel requires $L=64$ to achieve the same performance as the code with Arikan kernel and $L=256$. However, the code with the large kernel requires much lower number of iterations in average. Observe that decoding of both codes requires only summation and comparison operations.

\section{Conclusions}
In this memo performance and complexity results for polar subcodes and sequential successive cancellation decoding algorithm were presented. It was shown that optimized polar subcodes provide performance close to PAC\ codes of length 128, while extended BCH\ codes do outperform them.  The average complexity of sequential decoding of optimized polar subcodes is much less compared to PAC\ and extended BCH\ codes. It was shown that $(2048,1024)$ randomized polar subcode under sequential decoding provides both better performance and lower decoding complexity compared to the CCSDS LDPC\ code. Furthemore, $(4096,2048)$ polar subcode with the large kernel was shown to have lower sequential decoding complexity compared to the one with Arikan kernel.
\section*{Acknowledgements}
The author thanks G. Trofimiuk for providing an implementation of the decoder for polar codes with $K_2$ kernel, and N. Iakuba for his help with implementation of the sequential decoder.

\end{document}